\documentclass[12pt]{article}
\usepackage{float}
\usepackage{amssymb, amsmath, amsthm}
\usepackage{mathrsfs}
\usepackage{tabulary}
\usepackage{rotating}
\usepackage{enumerate}
\usepackage{setspace}
\usepackage{multirow}
\usepackage{paralist}
\usepackage{listings}
\usepackage[T1]{fontenc}
\usepackage[utf8]{inputenc}
\usepackage{authblk}
\usepackage{etoolbox}
\usepackage{longtable}
\usepackage[numbers,round,sort&compress]{natbib}
\usepackage[usenames,dvipsnames]{color}
\usepackage{bibentry}

\newcommand\BibTeX{{\rmfamily B\kern-.05em \textsc{i\kern-.025em b}\kern-.08em
T\kern-.1667em\lower.7ex\hbox{E}\kern-.125emX}}

\usepackage{hyperref}
\usepackage{subcaption}

\usepackage[hmargin=2.54cm,vmargin=2.54cm]{geometry}

\title{Robust and flexible estimation of data-dependent stochastic mediation effects: a proposed method and example in a randomized trial setting}
\author[1]{Kara E. Rudolph\thanks{Corresponding author: \\ 13B University Hall, Division of Epidemiology, School of Public Health, Berkeley, CA 94720\\
kara.rudolph@berkeley.edu \\ tel. +15107619404 }}
\author[2]{Oleg Sofrygin}
\author[3]{Wenjing Zheng}
\author[2]{Mark J. van der Laan}
\affil[1]{\footnotesize Division of Epidemiology, School of Public Health, University of California, Berkeley, California}
\affil[2]{\footnotesize Division of Biostatistics, School of Public Health, University of California, Berkeley, California}
\affil[3]{\footnotesize Center for AIDS Research, University of California, San Francisco, California}
\date{}

\begin{document}
\maketitle

\subsubsection*{Abstract}
Background: Causal mediation analysis can improve understanding of the mechanisms underlying epidemiologic associations. However, the utility of natural direct and indirect effect estimation has been limited by the assumption of no confounder of the mediator-outcome relationship that is affected by prior exposure---an assumption frequently violated in practice. \\

\noindent Methods: We build on recent work that identified alternative estimands that do not require this assumption and propose a flexible and double robust semiparametric targeted minimum loss-based estimator for data-dependent stochastic direct and indirect effects. The proposed method 
 treats the intermediate confounder affected by prior exposure as a time-varying confounder and intervenes stochastically on the mediator using a distribution which conditions on baseline covariates and marginalizes over the intermediate confounder. 
  In addition, we assume the stochastic intervention is given, conditional on observed data, which results in a simpler estimator and weaker identification assumptions.\\ 

\noindent Results: We demonstrate the estimator's finite sample and robustness properties in a simple simulation study. We apply the method to an example from the Moving to Opportunity experiment. In this application, randomization to receive a housing voucher is the treatment/instrument that influenced moving to a low-poverty neighborhood, which is the intermediate confounder. We estimate the data-dependent stochastic direct effect of randomization to the voucher group on adolescent marijuana use not mediated by change in school district and the stochastic indirect effect mediated by change in school district. We find no evidence of mediation. \\

\noindent Conclusions: Our estimator is easy to implement in standard statistical software, and we provide annotated R code to further lower implementation barriers. \\

\vspace{.2in}
\noindent
\textbf{Keywords: mediation; direct effect; indirect effect; double robust; targeted minimum loss-based estimation; targeted maximum likelihood estimation; data-dependent} 
\vspace{.2in}

\noindent

\newpage

\section{Introduction}

Mediation allows for an examination of the mechanisms driving a relationship. Much of epidemiology entails reporting exposure-outcome associations where the exposure may be multiple steps removed from the outcome. For example, risk-factor epidemiology demonstrated that obesity increases the risk of type 2 diabetes, but biochemical mediators linking the two have advanced our understanding of the causal relationship \citep{kahn2006mechanisms}. Mediators have been similarly important in understanding how social exposures act to affect health outcomes. In the illustrative example we consider in this paper, the Moving to Opportunity (MTO) experiment randomized families living in public housing to receive a voucher that they could use to rent housing on the private market, which reduced their exposure to neighborhood poverty \citep{kling2007experimental}. Ultimately, being randomized to receive a voucher affected subsequent adolescent drug use \citep{orr2003moving,rudolph2017composition}. In the illustrative example, we test the extent to which the effect operates through a change in the adolescent's school environment. 

Causal mediation analysis \citep{imai2010general,valeri2013mediation,zheng2012targeted} (also called mediation analysis using the counterfactual framework \citep{valeri2013mediation}) shares similar goals with the standard mediation approaches, e.g., structural equation modeling and the widely used Baron and Kenny ``product method'' approach \citep{baron1986moderator,valeri2013mediation}. They all aim to test mechanisms and estimate direct and indirect effects. Advantages of causal mediation analysis include that estimates have a causal interpretation (under specified identifying assumptions) and some approaches make fewer restrictive parametric modeling assumptions. For example, in contrast to traditional approaches, approaches within the causal mediation framework 1) allow for interaction between the treatment and mediator 
 \citep{VanderWeele2009marginal}, 
 2) allow for modeling nonlinear relationships between mediators and outcomes \citep{VanderWeele2009marginal}, 
 and 3) allow for incorporation of data-adaptive machine learning methods and double robust estimation \citep{zheng2012targeted,tchetgen2014estimation}. 

However, despite these advantages, the assumptions required to estimate certain causal mediation effects may sometimes be untenable; for example, the assumption that there is no confounder of the mediator-outcome relationship that is affected by treatment (in the literature, such a confounder is referred to as confounding by a causal intermediate \citep{petersen2006estimation}, a time-varying confounder affected by prior exposure \citep{vanderweele2016mediation}, or time-dependent confounding by an intermediate covariate \citep{van2012international}). For brevity, we will refer to such a variable as an intermediate confounder. 
  There have been recently proposed causal mediation estimands, called randomized (i.e., stochastic) interventional direct effects and interventional indirect effects that do not require this assumption \citep{didelez2006direct,van2008direct,zheng2012causal,VanderWeele2014effect,vansteelandt2017interventional,vanderweele2016mediation,zheng2017longitudinal}. We build on this work, proposing a robust and flexible estimator for these effects, 
 which we call stochastic direct and indirect effects (SDE and SIE).

This paper is organized as follows. In the following section, we review and compare common causal mediation estimands, providing the assumptions necessary for their identification. Then, we describe our proposed estimator, its motivation, and its implementation in detail. Code to implement this method is provided in the Appendix. We then provide results from a limited simulation study demonstrating the estimator's finite sample performance and robustness properties. Lastly, we apply the method in a longitudinal, randomized trial setting.

\section{Notation and Causal Mediation Estimands}

\begin{figure}[!h]
\caption{Structural causal model reflecting the illustrative example. }
\label{dag}
\centering
\includegraphics[width=\textwidth,height=0.7\textheight,keepaspectratio]{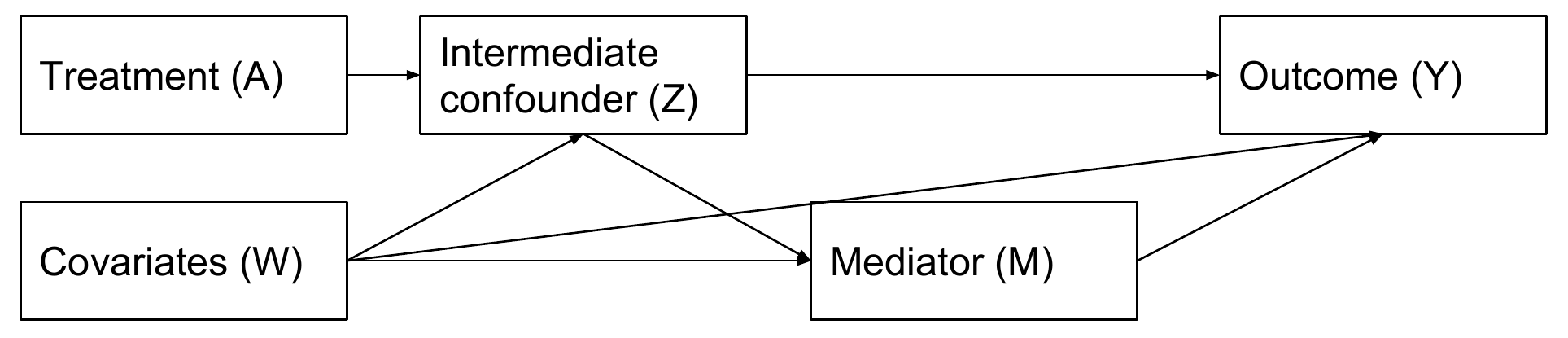}
\end{figure}

Let observed data: $O=(W, A, Z, M, Y)$ with $n$ i.i.d. copies $O_1,...,O_n \sim P_0$, where $W$ is a vector of pre-treatment covariates, $A$ is the treatment, $Z$ is the intermediate confounder affected by $A$, $M$ is the mediator, and $Y$ is the outcome. For simplicity, we assume that $A, Z, M, \text{ and }Y$ are binary. In our illustrative example, $A$ is an instrument, so it is reasonable to assume that $M$ and $Y$ are not affected by $A$ except through its effect on $Z$. Mirroring the structural causal model (SCM) of our illustrative example, we assume that $M$ is affected by $\{Z, W\}$ but not $A$, and that $Y$ is affected by $\{M, Z, W\}$ but not $A$. We assume exogenous random errors: ($U_W, U_A, U_Z, U_M, U_Y$). This SCM is represented in Figure \ref{dag} and the following causal models: $W=f(U_W)$, $A=f(U_A)$, $Z=f(A, W, U_Z)$, $M=f(Z,W,U_M)$, and $Y=f(M,Z,W,U_Y)$. Note that this SCM (including that $U_Y$ and $U_M$ are not affected by $A$) puts the following assumptions on the probability distribution: $P(Y | M, Z, A, W) = P(Y | M, Z, W)$ and $P(M | Z, A, W) = P(M | Z, W)$. However, our approach generalizes to scenarios where $A$ also affects $M$ and $Y$ as well as to scenarios where $A$ is not random. We provide details and discuss these generalizations in the Appendix.  
We can factorize the likelihood for the SCM reflecting our illustrative example as follows: $P(O) = P(Y | M, Z, W)P(M|Z,W)P(Z | A, W)P(A)P(W)$.

Causal mediation analysis typically involves estimating one of two types of estimands: controlled direct effects (CDE) or natural direct and indirect effects (NDE, NIE). Controlled direct effects involve comparing expected outcomes under different values of the treatment and setting the value of the mediator for everyone in the sample. For example the CDE can be defined: $E(Y_{a, m} - Y_{a*, m})$, where $Y_{a,m}$, $Y_{a^*,m}$, is the counterfactual outcome setting treatment $A$ equal to $a$ or $a^*$, respectively (the two treatment values being compared), and setting mediator $M$ equal to $m$. In contrast, the NDE can be defined: $E(Y_{a, M_{a^*}} - Y_{a*, M_{a^*}})$, where $Y_{a,M_{a^*}}$, $Y_{a^*,M_{a^*}}$ is the counterfactual outcome setting $A$ equal to $a$ or $a^*$ but this time setting $M_{a^*}$ to be the counterfactual value of the mediator had $A$ been set to $a^*$ (possibly contrary to fact). Similarly, the NIE can be defined: $E(Y_{a, M_a} - Y_{a, M_{a*}})$. Natural direct and indirect effects are frequently used in epidemiology and have the appealing property of adding to the total effect \citep{pearl2001direct}.

Although the NDE and NIE are popular estimands, their identification assumptions may sometimes be untenable. Broadly, identification of their causal effects relies on the sequential randomization assumption on intervention nodes $A$ and $M$ and positivity. 
 Two specific ignorability assumptions are required to identify CDEs and NDE/NIEs: 1) $A \perp Y_{a,m} | W$ and  2) $M \perp Y_{a,m} | W, A$ \citep{pearl2001direct}. The positivity assumptions are: $P(M =m | A=a,W) > 0 \text{  } a.e.$ and $P(A=a | W) > 0 \text{  } a.e.$ Two additional ignorability assumptions are required to identify NDE/NIEs: 3) $A \perp M_a | W$ and 4) $M_{a*} \perp Y_{a,m} | W$ \citep{pearl2001direct}. This last assumption states that, conditional on $W$, knowledge of $M$ in the absence of treatment $A$ provides no information of the effect of $A$ on $Y$ \citep{petersen2006estimation}. This assumption is violated when there is a confounder of the $M-Y$ relationship that is affected by $A$ (i.e., an intermediate confounder) \citep{avin2005identifiability,petersen2006estimation,vanderweele2016mediation}. This assumption is also problematic because it involves independence of counterfactuals under separate worlds ($a$ and $a^*$) which can never simultaneously exist. 

This last assumption that there is no confounder of the mediator-outcome relationship affected by prior treatment is especially concerning for epidemiology studies where longitudinal cohort data may reflect a data structure in which a treatment affects an individual characteristic measured at follow-up that in turn affects both a mediating variable and the outcome variable (see \citep{bild2002multi,eaton2012epidemiologic,phair1992acquired} for some examples). It is also problematic for mediation analyses involving instrumental variables such as randomized encouragement-design interventions where an instrument, $A$, encourages treatment uptake, $Z$, which then may influence $Y$ potentially through $M$. Such a design is present in the MTO experiment that we will use as an illustrative example. Randomization to receive a housing voucher ($A$) was the instrument that ``encouraged" the treatment uptake, moving with the voucher out of public housing ($Z$, which we will call the intermediate confounder). In turn, $Z$ may influence subsequent drug use among adolescent participants at follow-up ($Y$), possibly through a change the children's school environment ($M$). In the illustrative example, our goal was to examine mediation of the effect of receiving a housing voucher ($A$) on subsequent drug use ($Y$) by changing school districts ($M$) in the MTO data.

There has been recent work to relax the assumption of no intermediate confounder, $M_{a*} \perp Y_{a,m} | W$, by using a stochastic intervention on $M$ \citep{didelez2006direct,van2008direct,zheng2012causal,VanderWeele2014effect,vansteelandt2017interventional,vanderweele2016mediation,zheng2017longitudinal}. 
In this paper, we build on the approach described by 
 \cite{vanderweele2016mediation} in which they defined the stochastic distribution on $M$ as: $g_{M|a,W}$ or $g_{M|a^*,W}$, where \begin{equation} g_{M|A,W}(m,a^*,W) \equiv g_{M|a^*,W}(W) = \sum_{z=0}^1 P(M=1|Z=z,W)P(Z=z | A=a^*, W). \end{equation} (Equation 1 is an example of \cite{vanderweele2016mediation}'s work.) 
 In other words, instead of formulating the individual counterfactual values of $M_{a}$ or $M_{a^*}$, values are stochastically drawn from the distribution of $M$, conditional on covariates $W$ but marginal over intermediate confounder $Z$, setting $A=a$ or $A=a^*$, respectively. 
 The corresponding 
 estimands of interest are the $SDE = E(Y_{a,g_{M|a^*,W}}) - E(Y_{a*,g_{M|a^*,W}})$, and $SIE = E(Y_{a,g_{M|a,W}}) - E(Y_{a,g_{M|a^*,W}})$. 
 
 Others have taken a similar approach. For example, \cite{zheng2017longitudinal} formulate a stochastic intervention on $M$ that is fully conditional on the past: \begin{equation} \label{zhenggm} g_{M|Z,A,W}(m,Z,a^*,W) \equiv g_{M|Z,a^*,W} (Z,W) = P(M=1|Z, A=a^*, W) \end{equation} (note that per our SCM, $P(M=1 | Z,W) = P(M=1 | Z,A,W)$, so in our case, $g_{M|Z,a^*,W} (Z,W) = P(M=1|Z, W)$.) The corresponding estimands are the stochastic direct and indirect effects fully conditional on the past: $CSDE = E(Y_{a,g_{M|Z,a^*,W}}) - E(Y_{a*,g_{M|Z,a^*,W}})$, and $CSIE = E(Y_{a,g_{M|Z,a,W}}) - E(Y_{a,g_{M|Z,a^*,W}})$. However, \cite{zheng2017longitudinal}'s formulation shown in Equation \ref{zhenggm} is not useful for understanding mediation under the instrumental variable SCM we consider here, as there is no direct pathway from $A$ to $M$ or from $A$ to $Y$. Because of the restriction on our statistical model that $P(M | Z, A, W) = P(M | Z, W)$, $g_{M|Z,a^*,W} (Z,W) = g_{M|Z,a,W} (Z,W)$, so CSIE's under this model would equal 0. Thus, in this scenario, the NDE and CSDE are very different parameters. We note that it is also because of these restrictions on our statistical model stemming from the instrumental variable SCM that the sequential mediation analysis approach proposed by \cite{vanderweele2014mediation} would also result in indirect effects equal to 0. 

Because the CSIE and CSDE do not aid in understanding the role of $M$ as a potential mediator in this scenario, we focus instead on \cite{vanderweele2016mediation}'s SDE and SIE that condition on $W$ but marginalize over $Z$, thus completely blocking arrows into $M$ (similar to an NDE and NIE). The SDE and SIE coincide with the NDE and NIE in the absence of intermediate confounders \citep{vanderweele2016mediation}. 

\subsection{SDE and SIE Estimands and Identification} Our proposed estimator can be used to estimate two versions of the SDE and SIE: 1) fixed parameters that assume an unknown, true $g_{M|a^*,W}$; and 2)  data-dependent parameters that assume known $g_{M|a^*,W}$, estimated from the observed data, which we call $\hat{g}_{M|a^*,W}$. Researchers may have defensible reasons for choosing one version over the other, which we explain further below. The fixed SDE and SIE can be identified from the observed data distribution using the g-computation formula as discussed by \cite{vanderweele2016mediation}, assuming the sequential randomization assumption on intervention nodes $A$ and $M$: 1) $A \perp Y_{a,m} | W$, 2) $M \perp Y_{a,m} | W, A=a, Z $, and 3) $A \perp M_a | W$, for a particular $a$ and $g_{M|a^*,W}$. The data-dependent SDE And SIE can be identified similarly but need only the first two assumptions, 1) $A \perp Y_{a,m} | W$ and 2) $M \perp Y_{a,m} | W, A=a, Z $, because $\hat{g}_{M|a^*,W}$ is assumed known. If any of the above identification assumptions are violated, then the statistical estimands will not converge to their true causal quantities.

The estimand $E(Y_{a,\hat{g}_{M|a^*W}})$ can be identified via sequential regression, which provides the framework for our proposed estimator that follows. For intervention $(A=a, M=\hat{g}_{M|a^*,W})$, we have $\bar{Q}_M^{\hat{g}}(Z,W)\equiv \int_{m \in \mathcal{M}} \int_{y \in \mathcal{Y}} p(y | m, Z, W) d\mu_Y(y) \hat{g}_{m|a^*,W} d\mu_M(m)$, 
  where we integrate out $M$ under our stochastic intervention $\hat{g}_{M|a^*,W}$, and where $M$ has support $\mathcal{M}$ and $Y$ has support $\mathcal{Y}$ and where $d\mu_Y(y)$ and $d\mu_M(m)$ are some dominating measures. This is accomplished by evaluating $E(Y|M=m,Z=z,W)$ at each $m$ and multiplying it by the probability that $M=m$ under $\hat{g}_{M|a^*,W}$, summing over all $m$. We then integrate out $Z$ and set $A=a$: $\bar{Q}_Z^a(W)\equiv \int_{z \in \mathcal{Z}} \bar{Q}_M^{\hat{g}}(z,W|A=a,W) p(z | A=a, W) d\mu_Z(z)$, where $\mathcal{Z}$ denotes the support of random variable $Z$ and $d\mu_Z(z)$ is some dominating measure. Marginalizing over the distribution of $W$ gives the statistical parameter: $\Psi(P)(a,\hat{g}_{M|a^*,W})= \int_{w \in \mathcal{W}} \bar{Q}_Z^a(w) p(w) d\mu_W(w)$, where $\mathcal{W}$ denotes the support of random variable $W$ and $d\mu_W(w)$ is some dominating measure.

In the next section, we propose a novel, robust substitution estimator that can be used to estimate both the fixed and parametric versions of the SDE and SIE. Inference for the fixed SDE and SIE can be obtained by using the bootstrapped variance, which requires parametric models for the nuisance parameters $P(A)$ and $P(M | Z,A,W)$. This is the same inference strategy as proposed by \cite{vanderweele2016mediation}. However, researchers may encounter scenarios for which fitting parametric models is unappealing and using machine learning approaches is preferred. The data-dependent SDE and SIE with inference based on the efficient influence curve (EIC) may be preferable in such scenarios. In contrast to the EIC for an assumed known $g_{M|a^*,W}$, the EIC an unknown $g_{M|a^*,W}$ is complicated due to the dependence of the unknown marginal stochastic intervention 
 on the data distribution. Such an EIC would include an $M$ component, the form of which would be more complex due to the distribution of $M$ being marginalized over $Z$. No statistical tools for solving an EIC of that form currently exist. Solving the EIC for the the parameter $\Psi(P)(a,g_{M|a^*,W})$ for an unknown $g_{M|a^*,W}$ is ongoing work.

\section{Targeted minimum loss-based estimator}


Our proposed estimator 
 uses targeted minimum loss-based estimation (TMLE) \citep{van2006targeted}, targeting the stochastic, counterfactual outcomes that comprise the SDE and SIE. To our knowledge, it is the first such estimator appropriate for instrumental variable scenarios. 
 TMLE is a substitution estimation method that solves the EIC estimating equation. Its robustness properties differ for the fixed and data-dependent parameters. For the data-dependent SDE and SIE, if either the $Y$ model is correct or the $A$ and $M$ models given the past are correct, then one obtains a consistent estimator of the parameter. Robustness to misspecification of the treatment model is relevant under an SCM with nonrandom treatment; we discuss the generalization of our proposed estimator to such an SCM in the Appendix. Note that $\hat{g}_{M|a^*,W}$ for the stochastic intervention is not the same as the conditional distribution of $M$ given the past, so the first could be inconsistent while the latter is consistent. For the fixed SDE and SIE, we also need to assume consistent estimation of $g_{M|a^*,W}$, since it does not target $g_{M|a^*,W}$ (and is therefore not a full TMLE for the fixed parameters).  

The estimator integrates two previously developed TMLEs: one for stochastic interventions \citep{munoz2012population} and one for multiple time-point interventions \citep{van2012international}, which is built on the iterative/recursive g-computation approach \citep{bang2005doubly}. This TMLE is not efficient under the SCM considered, because of the restriction on our statistical model that $P(Y | M,Z,A,W) = P(Y|M,Z,W)$. However, it is still a consistent estimator if that restriction on our model does not hold (i.e., $P(Y | M,Z,A,W) \ne P(Y|M,Z,W)$), because the targeting step adds dependence on $A$.
The TMLE is constructed using the sequential regressions described in the above section with an additional targeting step after each regression. The TMLE solves the EIC for the target parameter that treats $g_{M|a^*,W}$ as given. A similar EIC has been described previously \citep{bang2005doubly,van2012international}. The EIC for the parameter $\Psi(P)(a, g_{M|a^*,W})$ for a given $g_{M|a^*,W}$ is given by:
\begin{equation} \begin{split}
    D^*(a, g_{M|a^*,W}) & = \sum_{k=0}^2 D_k^*(a, g_{M|a^*,W}), \text{ where }\\
D^*_0(a, g_{M|a^*,W}) & = \bar{Q}^a_Z(W) - \Psi(P)(a, g_{M|a^*,W})\\
    D^*_1(a, g_{M|a^*,W}) & = \frac{I(A=a)}{P(A=a | W)}(\bar{Q}^{g}_M(Z,W) - \bar{Q}^a_Z(W))\\
    D^*_2(a, g_{M|a^*,W}) & = \frac{I(A=a)\{I(M=1)g_{M|a^*,W} + I(M=0)(1-g_{M|a^*,W}) \}}{P(A=a | W)\{I(M=1)g_{M|z,W} + I(M=0)(1-g_{M|z,W}) \}}\\
    & \times (Y-\bar{Q}_Y(M,Z,W)).   
\end{split}
\end{equation}
Substitution of $g_{M|a^*,W}=\hat{g}_{M|a^*,W}$ yields the EIC used for the data-dependent parameter $\Psi(P)(a, \hat{g}_{M|a^*,W})$. The EIC for the parameter $\Psi(P)(a, g_{M|a^*,W})$ in which the stochastic intervention equals the unknown $g_{M|a^*,W}$ is an area of future work.

We now describe how to compute the TMLE. In doing so, we use parametric model/regression language for simplicity but data-adaptive estimation approaches that incorporate machine learning \citep[e.g.,][]{van2007super} may be substituted and may be preferable (we use such a data-adaptive approach in the illustrative example analysis). We note that survey or censoring weights could be incorporated into this estimator as described previously \citep{rudolph2014estimating}. We use notation reflecting estimation of the data-dependent parameters, but note that estimation of the fixed parameters would be identical---in the fixed parameter case, the notation would refer to $g_{M|a^*,W}$ instead of $\hat{g}_{M|a^*,W}$.  

First, one estimates $\hat{g}_{M|a^*,W}(W)$, which is the estimate of $g_{M|a^*,W}(W)$, defined in Equation 1, using observed data. Consider a binary $Z$. We estimate $g_{Z|a^*,W}(W)=P(Z=1 | A=a^*, W)$. We then estimate $g_{M|z,W}(W)=P(M=1 | Z=z, W)$ for $z \in \{0,1\}$. We use these quantities to calculate $\hat{g}_{M|a^*,W} = \hat{g}_{M|z=1,W}\hat{g}_{Z|a^*,W} +  \hat{g}_{M|z=0,W}(1-\hat{g}_{Z|a^*,W})$. We can obtain $\hat{g}_{Z|a^*,W}(W)$ from a logistic regression of $Z$ on $A, W$ setting $A=a^*$, and $\hat{g}_{M|z,W}(W)$ from a logistic regression of $M$ on $Z, W$, setting $Z=\{0,1\}$. We will then use this stochastic intervention in the TMLE, whose implementation is described as follows.

\begin{enumerate}
\item Let $\bar{Q}_{Y,n}(M,Z,W)$ be an estimate of $\bar{Q}_Y(M,Z,W)\equiv E(Y |M,Z,W)$. To obtain $\bar{Q}_{Y,n}(M,Z,W)$, predict values of $Y$ from a regression of $Y$ on $M,Z,W$. 
\item Estimate the weights to be used for the initial targeting step:\\ $h_1(a) = \frac{I(A=a)\{I(M=1)\hat{g}_{M|a^*,W} + I(M=0)(1-\hat{g}_{M|a^*,W}) \}}{P(A=a)\{I(M=1)g_{M|Z,W} + I(M=0)(1-g_{M|Z,W}) \}},$ where estimates of $g_{M|Z,W}$ are predicted probabilities from a logistic regression of $M=m$ on $Z$ and $W$. Let $\hat{h}_{1,n}(a)$ denote the estimate of $h_1(a)$. 
\item Target the estimate of $\bar{Q}_{Y,n}(M,Z,W)$ by considering a univariate parametric submodel $\{\bar{Q}_{Y,n}(M,Z,W)(\epsilon):\epsilon\}$ defined as:
$logit(\bar{Q}_{Y,n} (\epsilon)(M,Z,W)) = logit(\bar{Q}_{Y,n}(M,Z,W) ) + \epsilon$. 
 Let $\epsilon_n$ be the MLE fit of $\epsilon$. We obtain $\epsilon_n$ by setting $\epsilon$ as the intercept of a weighted logistic regression model of $Y$ with $logit(\bar{Q}_{Y,n}(M,Z,W))$ as an offset and weights $\hat{h}_{1,n}(a)$. (Note that this is just one possible TMLE.)
 The update is given by $ \bar{Q}^*_{Y,n}(M,Z,W) =  \bar{Q}_{Y,n}(\epsilon_n)(M,Z,W).$ $Y$ can be bounded to the [0,1] scale as previously recommended \citep{gruber2010targeted}.
\item Let $\bar{Q}^{g}_{M,n}(Z,W)$ be an estimate of $\bar{Q}^{g}_{M}(Z,W)$. To obtain $\bar{Q}^{g}_{M,n}(Z,W)$, we integrate out $M$ to from $\bar{Q}^*_{Y,n}(M,Z,W)$. First, we estimate $\bar{Q}^*_{Y,n}(M,Z,W)$ setting $m=1$ and $m=0$, giving $\bar{Q}^*_Y(m=1, z, w)$ and $\bar{Q}^*_Y(m=0, z, w)$. 
 Then, multiply these predicted values by their probabilities under 
 $\hat{g}_{M|a^*,W}(W)$ (for $a \in \{a, a^{*}\}$), and add them together (i.e., $\bar{Q}^{\hat{g}}_{M,n}(Z,W)= \hat{Q}^*_Y(m=1, z, w)\hat{g}_{M|a^*,W} + \hat{Q}^*_Y(m=0, z, w)(1-\hat{g}_{M|a^*,W})$).
\item 
 We now fit a regression of $\bar{Q}^{\hat{g},*}_{M,n}(Z,W)$ on $W$ among those with $A=a$. We call the predicted values from this regression $\bar{Q}^{a}_{Z,n}(W)$. The empirical mean of these predicted values is the TMLE estimate of $\Psi(P)(a, \hat{g}_{M|a^*,W})$.
\item Repeat the above steps for each of the interventions. For example, for binary $A$, we would execute these steps a total of three times to estimate: 1) $\Psi(P)(1,\hat{g}_{M|1,W})$, 2) $\Psi(P)(1,\hat{g}_{M|0W})$, and 3) $\Psi(P)(0,\hat{g}_{M|0,W})$. 
\item The SDE can then be obtained by substituting estimates of parameters $\Psi(P)(a,\hat{g}_{M|a^*,W}) - \Psi(P)(a^*,\hat{g}_{M|a^*,W})$ and the SIE can be obtained by substituting estimates of parameters $\Psi(P)(a,\hat{g}_{M|a,W}) - \Psi(P)(a,\hat{g}_{M|a^*,W})$.
\item For the fixed parameters, the variance can be estimated using the bootstrap. For the data-dependent parameters, the variance of each estimate from Step 6 can be estimated as the sample variance of the EIC (defined above, substituting in the targeted fits $\bar{Q}^*_{Y,n}(M,Z,W)$ and $\bar{Q}^{a,*}_{Z,n}(W)$) divided by $n$. First, we estimate the EIC for each component of the data-dependent SDE/SIE, which we call $EIC_{\Psi(P)(a,\hat{g}_{M|a^*,W})}$. 
 Then we estimate the EIC for the estimand of interest by subtracting the EICs corresponding to the components of the estimand. For example $EIC_{SDE} = EIC_{\Psi(P)(a,\hat{g}_{M|a^*,W})} - EIC_{\Psi(P)(a^*,\hat{g}_{M|a^*,W})}$. The sample variance of this EIC divided by $n$ is the influence curve (IC)-based variance of the data-dependent estimator. 
\end{enumerate}

\section{Simulation}
\subsection{Data generating mechanism}
We conduct a simulation study to examine finite sample performance of the TMLE estimators for the fixed SDE and SIE and data-dependent SDE and SIE from the data-generating mechanism (DGM) shown in Table \ref{dgmtab}. Under this DGM, the data-dependent SDE is: $SDE=E(Y_{1, \hat{g}_{M|0,W}}) - E(Y_{0, \hat{g}_{M|0,W}})$ and the SIE is: $SIE=E(Y_{1, \hat{g}_{M|1,W}}) - E(Y_{1, \hat{g}_{M|0,W}})$. The fixed versions are defined with respect to the unknown, true $g_{M|1,W}$ and $g_{M|0,W}$. Table \ref{dgmtab} uses the same notation and SCM as in Section 2, with the addition of $\Delta$, an indicator of selection into the sample (which corresponds to the MTO data used in the empirical illustration where one child from each family is selected to participate). 

\begin{table}[!h]
\centering
\caption{Simulation data-generating mechanism .}
\label{dgmtab}
\begin{tabular}{| p{10cm} | p{3cm} | }
  \hline
$W_1 \sim Ber(0.5)$ & $P(W_1=1)=0.50$ \\
$W_2 \sim Ber(0.4 + 0.2W_1)$  & $P(W_2=1)=0.50$\\
$\Delta \sim Ber(-1 + log(4)W_1 + log(4)W_2$  & $P(\Delta =1)=0.58$\\
$A = \Delta A^*$, where $A^*\sim Ber(0.5)$  & $P(A=1)=0.50$ \\
$Z = \Delta Z^*$, where $Z^*\sim Ber(log(4)A - log(2)W_2) $ & $P(Z=1)=0.58$\\
$M = \Delta M^*$, where $M^*\sim Ber(-log(3) + log(10)Z - log(1.4)W_2)$ & $P(M=1)=0.52$\\
$Y = \Delta Y^*$, where $Y^*\sim Ber(log(1.2) + log(3)Z + log(3)M - log(1.2)W_2 + log(1.2)ZW_2)$ & $P(Y=1)=0.76$\\
   \hline
\end{tabular}
\end{table}

We compare performance of the TMLE estimator to an inverse-probability weighted estimator (IPTW) and estimator that solves the EIC estimating equation (EE) but differs from TMLE in that it lacks the targeting steps and is not a plug-in estimator, so its estimates are not guaranteed to lie within the parameter space (which may be particularly relevant for small sample sizes). Variance for the fixed SDE and SIE parameters is calculated using 500 bootstrapped samples for each simulation iteration. Variance for the data-dependent SDE and SIE is calculated using the EIC. We show estimator performance in terms of absolute bias, percent bias, closeness to the efficiency bound (mean estimator standard error (SE) $\times$ the square root of the number of observations), 95\% confidence interval (CI) coverage, and mean squared error (MSE) across 1,000 simulations for sample sizes of N=5,000, N=500, and N=100. In addition, we consider 1) correct specification of all models, and 2) misspecification of the $Y$ model that included a term for $Z$ only. 

\subsection{Performance}
Table \ref{restabfixed} gives simulation results under correct model specification for fixed SDE and SIE using bootstrap-based variance. Table \ref{restabdatadependent} gives simulation results under correct model specification for data-dependent SDE and SIE using IC-based variance. Both Tables \ref{restabfixed} and \ref{restabdatadependent} show that the TMLE, IPTW, and EE estimators are consistent when all models are correctly specified, showing biases of around 1\% or less under large sample size (N=5,000) and slightly larger biases with smaller sample sizes. The 95\% CIs for the TMLE and EE estimators result in similar coverage that is close to 95\%, except for estimation of the SIE with a sample size of N=100, which has coverage closer to 90\%.
Confidence intervals for the IPTW estimator for the fixed parameter are close to 95\% but are conservative and close to 100\% for the data-dependent parameter. As expected, IPTW is less efficient than TMLE or EE; the TMLE and EE estimators perform similarly and close to the efficiency bound for all sample sizes.

\begin{table}[H]
\centering
\caption{Simulation results for fixed SDE and SIE using bootstrapped-based variance (500 boostrapped samples) under correct specification of all parametric models for various sample sizes. 1,000 simulations. Estimation methods compared include targeted minimum loss-based estimation (TMLE), inverse probability weighting estimation (IPTW), and solving the estimating equation (EE). Bias and MSE values are averages across the simulations. The estimator standard error $\times \sqrt{n}$ should be compared to the efficiency bound, which is 1.07 for the SDE and 0.24 for the SIE.}
\label{restabfixed}
\begin{tabular}{|p{2cm}| p{1.8cm}  p{1.5cm} p{2cm}  p{2.5cm}  p{1.8cm}  | }
  \hline
Estimand & Bias & \%Bias & SE$\times \sqrt{n}$ & 95\%CI Cov & MSE \\   \hline 
\multicolumn{6}{| c | }{ N=5,000} \\ \hline
TMLE&&&&&\\
SDE & -3.95e-04 & -0.58 & 1.11 & 94.1 & 2.50e-04 \\ 
SIE & 5.51e-05 & 0.17 & 0.25 & 94.9 & 1.29e-05 \\ 
\hline
IPTW&&&&&\\
SDE  &5.29e-04&0.78&1.69&95.2&3.39e-04  \\ 
SIE &8.96e-06&0.03&0.42&94.5&5.42e-04 \\ 
\hline
EE&&&&&\\
SDE&7.90e-04&1.17&1.11&93.7&2.76e-04 \\ 
SIE&2.53e-04&0.96&0.25&94.8&1.22e-05\\
\hline
\multicolumn{6}{| c | }{ N=500} \\ \hline
TMLE&&&&&\\
SDE & 1.87e-04 & 0.27 & 1.31 & 94.4 & 2.46e-03  \\ 
SIE & -2.91e-04 & -1.10 & 0.41 & 94.7 & 7.89e-04  \\ 
\hline
IPTW&&&&&\\
SDE &-2.79e-03 &-4.12 &1.68 & 94.7& 5.60e-03\\ 
SIE &1.44e-03 &5.45& 0.44 &95.0&3.64e-04\\ 
\hline
EE&&&&&\\
SDE  &-1.69e-03 & -2.49& 1.10&94.0 &2.49e-03\\ 
SIE  & 3.22e-04 & 1.22& 0.26&94.0& 1.27e-04\\ 
\hline
\multicolumn{6}{| c | }{ N=100} \\ \hline
TMLE&&&&&\\
SDE &6.78e-03 & 10.02 & 1.09 & 98.4 & 1.36e-02 \\ 
SIE & -1.58e-03 & -5.98 & 0.25 & 87.9 & 1.26e-04 \\ 
\hline
IPTW&&&&&\\
SDE  & -3.20e-03 &-4.73 & 1.68 &94.1& 2.96e-02 \\ 
SIE &-5.77e-04&-2.18& 0.53 &95.2&2.02e-03\\ 
\hline
EE&&&&&\\
SDE  &  2.83e-03 & 4.18 & 1.09& 94.9& 1.19e-02 \\ 
SIE  &  -5.25e-04& -1.98& 0.31&93.0&7.49e-04\\ 
\hline
\end{tabular}
\end{table}

\begin{table}[H]
\centering
\caption{Simulation results for data-dependent SDE and SIE using influence curve-based variance under correct specification of all parametric models for various sample sizes. 1,000 simulations. Estimation methods compared include targeted minimum loss-based estimation (TMLE), inverse probability weighting estimation (IPTW), and solving the estimating equation (EE). Bias and MSE values are averages across the simulations. The estimator standard error $\times \sqrt{n}$ should be compared to the efficiency bound, which is 1.07 for the SDE and 0.24 for the SIE.}
\label{restabdatadependent}
\begin{tabular}{|p{2cm}| p{1.8cm}  p{1.5cm} p{2cm}  p{2.5cm}  p{1.8cm}  | }
  \hline
Estimand & Bias & \%Bias & SE$\times \sqrt{n}$ & 95\%CI Cov & MSE \\   \hline 
\multicolumn{6}{| c | }{ N=5,000} \\ \hline
TMLE&&&&&\\
SDE & 1.08e-03 & 1.61 & 1.11 & 93.09 & 2.77e-04 \\ 
SIE &  8.21e-06 & 0.03 & 0.24 & 94.89 & 1.10e-05 \\ 
\hline
IPTW&&&&&\\
SDE& 7.87e-04 & 1.12 & 2.28 & 99.40 & 6.16e-04 \\ 
SIE & 6.51e-06 & 0.05 & 1.18 & 100.00 & 3.74e-05 \\ 
\hline
EE&&&&&\\
SDE&1.20e-03 & 1.79 & 1.12 & 93.71 & 2.76e-04 \\ 
SIE & 1.85e-05 & 0.06 & 0.24 & 95.21 & 1.09e-05 \\   
\hline
\multicolumn{6}{| c | }{ N=500} \\ \hline
TMLE&&&&&\\
SDE& 7.55e-04 & 1.14 & 1.10 & 95.50 & 2.29e-03 \\ 
SIE & -4.33e-04 & -1.36 & 0.23 & 94.59 & 1.20e-04 \\ 
\hline
IPTW&&&&&\\
SDE& 6.28e-03 & 9.00 & 2.29 & 98.80 & 5.75e-03 \\ 
SIE & -1.90e-03 & -6.44 & 1.19 & 100.00 & 3.76e-04 \\ 
\hline
EE&&&&&\\
SDE & 8.27e-04 & 1.24 & 1.11 & 95.51 & 2.32e-03 \\ 
SIE & -3.35e-04 & -0.92 & 0.24 & 94.31 & 1.24e-04 \\ 
\hline
\multicolumn{6}{| c | }{ N=100} \\ \hline
TMLE&&&&&\\
SDE& 6.34e-03 & 8.81 & 1.07 & 95.50 & 1.30e-02 \\ 
SIE & -1.90e-03 & -7.85 & 0.21 & 87.99 & 7.45e-04 \\ 
\hline
IPTW&&&&&\\
SDE& -1.07e-02 & -16.87 & 2.31 & 99.40 & 2.84e-02 \\ 
SIE & -1.35e-03 & -5.91 & 1.18 & 100.00 & 2.29e-03 \\
\hline
EE&&&&&\\
SDE& 1.29e-03 & 1.27 & 1.10 & 97.01 & 1.21e-02 \\ 
SIE & 2.44e-04 & 0.15 & 0.23 & 90.12 & 7.94e-04 \\ 
\hline
\end{tabular}
\end{table}

Table \ref{restabfixedmissy} gives simulation results under misspecification of the outcome model that only includes a term for $Z$ for fixed SDE and SIE using bootstrap-based variance. Thus, comparing results in Table \ref{restabfixedmissy} to Table \ref{restabfixed} demonstrates robustness to misspecification of the outcome model. As all three of the estimators evaluated are theoretically robust to misspecification of this model, we would expect similar results between the two Tables, and we see that is indeed the case.

\begin{table}[H]
\centering
\caption{Simulation results for fixed SDE and SIE using bootstrapped-based variance (500 bootstrapped samples) under misspecification of the outcome model for various sample sizes. 1,000 simulations. Estimation methods compared include targeted minimum loss-based estimation (TMLE), inverse probability weighting estimation (IPTW), and solving the estimating equation (EE). Bias and MSE values are averages across the simulations. The estimator standard error $\times \sqrt{n}$ should be compared to the efficiency bound, which is 1.07 for the SDE and 0.24 for the SIE.}
\label{restabfixedmissy}
\begin{tabular}{|p{2cm}| p{1.8cm}  p{1.5cm} p{2cm}  p{2.5cm}  p{1.8cm}  | }
  \hline
Estimand & Bias & \%Bias & SE$\times \sqrt{n}$ & 95\%CI Cov & MSE \\   \hline 
\multicolumn{6}{| c | }{ N=5,000} \\ \hline
TMLE&&&&&\\
SDE &-2.21e-03&0.16&1.13&95.3& 2.38e-04 \\ 
SIE &1.79e-04&0.678&0.25&95.5& 1.23e-05\\ 
\hline
IPTW&&&&&\\
SDE  &-1.25e-03& -1.85&1.68&93.8&5.76e-04  \\ 
SIE & 4.56e-04&1.72& 0.42&93.7& 3.59e-05 \\ 
\hline
EE&&&&&\\
SDE&-1.81e-04&-0.27&1.13&94.8& 2.36e-04\\ 
SIE&3.07e-04&1.16&0.26&94.3& 1.34e-05\\
\hline
\multicolumn{6}{| c | }{ N=500} \\ \hline
TMLE&&&&&\\
SDE & 1.58e-03& 2.33& 1.12 & 94.6 & 2.49e-03\\ 
SIE & 7.00e-05 & 0.26& 0.26 & 95.1&1.30e-04 \\ 
\hline
IPTW&&&&&\\
SDE & -3.08e-03& -4.55& 1.70& 94.8&5.74e-03\\ 
SIE &2.25e-04& 0.85& 0.44& 94.5&3.86e-04\\ 
\hline
EE&&&&&\\
SDE  &1.86e-03 & 2.75& 1.12 & 93.1& 2.63e-03\\ 
SIE  & -4.24e-04&-1.60& 0.26 & 94.1&1.28e-04\\ 
\hline
\multicolumn{6}{| c | }{ N=100} \\ \hline
TMLE&&&&&\\
SDE & 6.93e-04 & 1.02&1.11 & 93.4& 1.33e-02\\ 
SIE & -7.91e-04& -2.99&0.28& 90.1 & 6.99e-04\\ 
\hline
IPTW&&&&&\\
SDE  &6.76e-03&9.99& 1.77& 94.4&3.05e-02\\ 
SIE & -2.21e-03& -8.37&0.55& 94.7& 2.09e-03\\ 
\hline
EE&&&&&\\
SDE  &4.81e-03 & 7.10& 1.16&93.6& 1.28e-02 \\ 
SIE  &1.09e-03& 4.12& 0.34 & 93.1& 8.85e-04  \\ 
\hline
\end{tabular}
\end{table}


\section{Empirical Illustration}
\subsection{Overview and set-up}
We now apply our proposed estimator to MTO: a longitudinal, randomized trial that is described above. Because we wish to use machine learning for this empirical illustration, we will estimate the data-dependent SDE of being randomized to receive a housing voucher ($A$) on marijuana use ($Y$) not mediated by change in school district ($M$) and the data-dependent SIE mediated by $M$ among adolescent boys in the Boston site in the presence of an intermediate confounder ($Z$), moving with the voucher out of public housing. 

We restrict to adolescents less than 18 years old who were present at interim follow-up, as those participants had school data and were eligible to be asked about marijuana use. We restrict to boys in the Boston site as previous work has shown important quantitative and qualitative differences in MTO's effects by sex \citep{sanbonmatsu2011moving,kling2005neighborhood,orr2003moving,leventhal2011moving,osypuk2012differential,osypuk2012gender} and by city \citep{rudolph2017composition}. We choose to present results from a restricted analysis instead of a stratified analysis, as our goal is to illustrate the proposed method. A more thorough mediation analysis considering all sexes and sites is the subject of a future paper. 

Marijuana use was self-reported by adolescents at the interim follow-up, which occurred 4-7 years after baseline, and is defined as ever versus never use. Change in school district is defined as the school at follow-up and school at randomization being in the same district. Numerous baseline characteristics included individual and family sociodemographics, motivation for participating in the study, neighborhood perceptions, school-related characteristics of the adolescent, and predictive interactions. 

We use machine learning to flexibly and data-adaptively model the following relationships: instrument to intermediate confounder, intermediate confounder to mediator, and mediator to outcome. Specifically, we use least absolute shrinkage and selection operator (lasso) \citep{tibshirani1996regression} and choose the model that improves 10-fold cross-validation prediction error, while always including age and race/ethnicity and relevant $A, Z,$ and $M$ variables. 

\subsection{Results}
\begin{figure}[!h]
\caption{Mediated effect estimates and 95\% confidence intervals using interim follow-up data from adolescent boys in the Boston site of the Moving to Opportunity experiment. The data-dependent SDE is interpreted as the direct effect of being randomized to receive a housing voucher on risk of marijuana use that is not mediated through a change in school district. The data-dependent SIE is interpreted as the effect of being randomized to receive a housing voucher on marijuana use that is mediated by changing school districts. }
\label{mtoresfig}
\centering
\includegraphics[width=.8\textwidth,keepaspectratio]{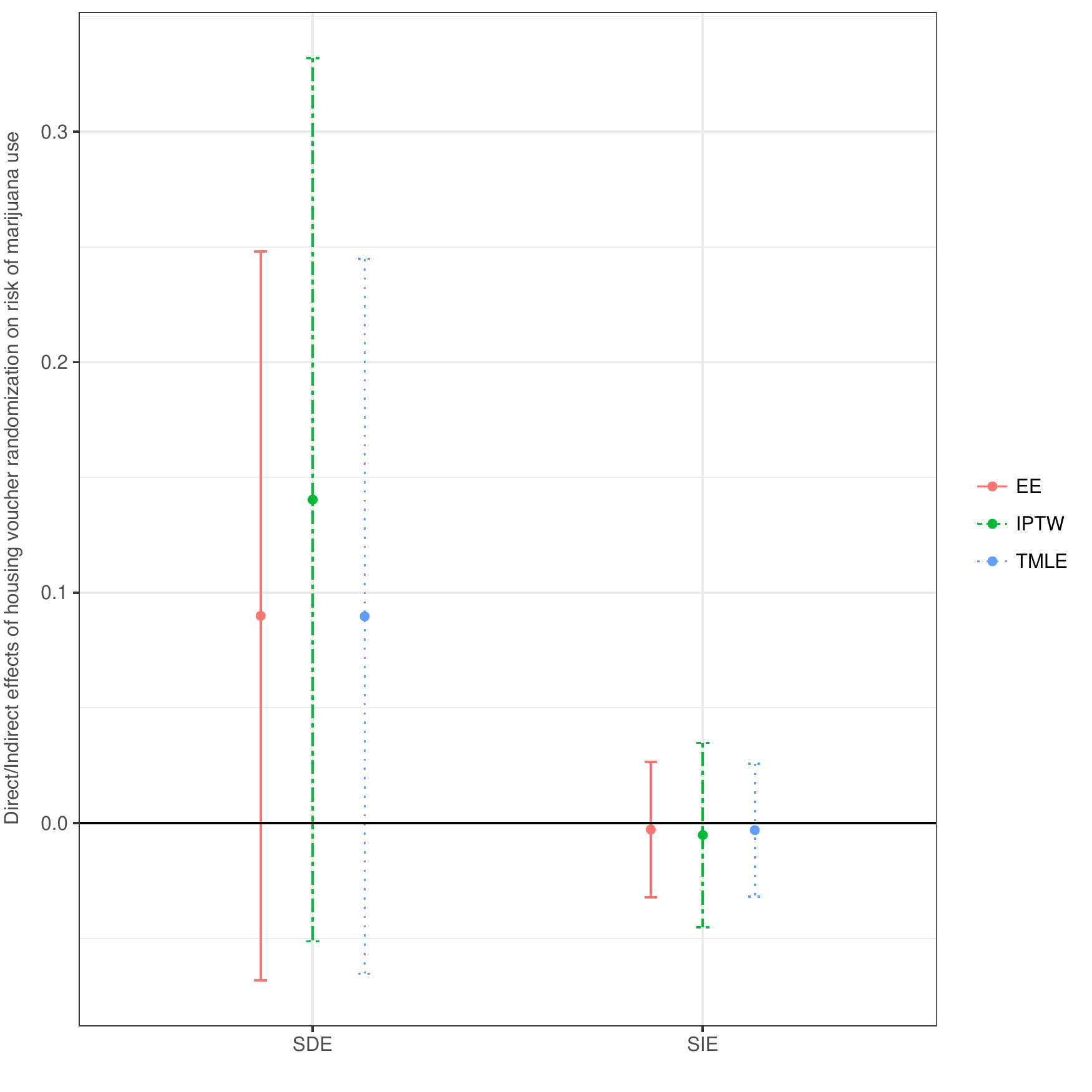}
\end{figure}

Figure \ref{mtoresfig} shows the data-dependent SDE and SIE estimates by type of estimator (TMLE, IPTW, and EE) for boys in the Boston MTO site (N=228). SDE and SIE estimates are similar across estimators. We find no evidence that change in school district mediated the effect of being randomized to the voucher group on marijuana use, with null SIE estimates (TMLE risk difference: -0.003, 95\% CI: -0.032, 0.026). The direct effect of randomization to the housing voucher group on marijuana use suggests that boys who were randomized to this group were 9\% more likely to use marijuana than boys in the control group, though this difference is not statistically significant (TMLE risk difference: 0.090, 95\% CI: -0.065-0.245). 

\section{Discussion}
We proposed robust targeted minimum loss-based estimators to estimate fixed and data-dependent stochastic direct and indirect effects that are the first to naturally accommodate instrumental variable scenarios. 
 These estimators build on previous work identifying and estimating the SDE and SIE \citep{vanderweele2016mediation}. The SDE and SIE have the appealing properties of 1) relaxing the assumption of no intermediate confounder affected by prior exposure, and 2) utility in studying mediation in the context of instrumental variables that adhere to the exclusion restriction assumption (a common assumption of instrumental variables which states that there is no direct effect between $A$ and $Y$ or between $A$ and $M$ \citep{angrist1996identification}) due to completely blocking arrows into the mediator by marginalizing over the intermediate confounder, $Z$. Given the restrictions that this assumption places on the statistical model, several alternative estimands are not appropriate for understanding mediation in this context as the indirect effect would always equal zero \citep[e.g.,][]{zheng2017longitudinal,tchetgen2013inverse,vanderweele2014mediation}. 
 
Inference for the fixed SDE and SIE can be obtained from bootstrapping, using parametric models for nuisance parameters. Inference for the data-dependent SDE and SIE can be obtained from the data-dependent EIC that assumes known $\hat{g}_{M|a^*,W}$ estimated from the data, and is appropriate for integrating machine learning in modeling nuisance parameters. The ability to incorporate machine learning is a significant strength in this case; if using the parametric alternative, multiple models would need to be correctly specified \citep{vanderweele2016mediation}.  
IC-based variance is possible in estimating the data-dependent SDE and SIE, because the data-dependent EIC has a form that is solvable using existing statistical tools; in contrast, the EIC for the fixed parameters is more complex and is not solvable with current statistical tools. 

Our proposed estimator for the fixed and data-dependent parameters is simple to implement in standard statistical software, and we provide R code to lower implementation barriers. 
 Another advantage of our TMLE estimator, which is shared with other estimating equation approaches, is that it is robust to some model misspecification. In estimating the data-dependent SDE and SIE, one could obtain a consistent estimate as long as either the $Y$ model or the $A$ and $M$ models given the past were correctly specified. Obtaining a consistent estimate of the fixed SDE and SIE would also require consistent estimation of $g_{M|a^*,W}$. 
  In addition, our proposed estimation strategy is less sensitive to positivity violations than weighting-based approaches. First, TMLE is usually less sensitive to these violations than weighting estimators, due in part to it being a substitution estimator, which means that its estimates lie within the global constraints of the statistical model. This is in contrast to alternative estimating equation approaches, which may result in estimates that lie outside the parameter space. Second, we formulate our TMLE such that the targeting is done as a weighted regression, which may smooth highly variable weights \citep{stitelman2012general}. In addition, moving the targeting into the weights 
 improves computation time \citep{stitelman2012general}.

However, there are also limitations to the proposed approach. We have currently only implemented it for a binary $A$ and $M$, though extensions to multinomial or continuous versions of those variables are possible \citep{rosenblum2010targeted,diaz2015targeted}. Extending the estimator to allow for a high-dimensional $M$ is less straightforward, though it is of interest and an area for future work as allowing for high-dimensional $M$ is a strength of other mediation approaches \citep{tchetgen2013inverse,zheng2017longitudinal}. We also plan to focus future work on developing a full TMLE for the fixed SDE and SIE parameters. 

\bibliographystyle{DeGruyter}
\bibliography{mediation}

\appendix

\section{Generalizations to other structural causal models}
\subsection{Nonrandom Treatment}
Let observed data: $O=(W, A, Z, M, Y)$ with $n$ i.i.d. copies $O_1,...,O_n \sim P_0$, where $W$ is a vector of pre-treatment covariates, $A$ is the treatment, $Z$ is the intermediate confounder affected by $A$, $M$ is the mediator, and $Y$ is the outcome. For simplicity, we assume that $A, Z, M, \text{ and }Y$ are binary. We assume that $M$ and $Y$ are not affected by $A$ except through its effect on $Z$. We assume that $A$ is affected by $\{W\}$, $Z$ is affected by $\{A, W\}$, $M$ is affected by $\{Z, W\}$ but not $A$, and that $Y$ is affected by $\{M, Z, W\}$ but not $A$. We assume exogenous random errors: ($U_W, U_A, U_Z, U_M, U_Y$). Note that this SCM (including that $U_Y$ and $U_M$ are not affected by $A$) puts the following assumptions on the probability distribution: $P(Y | M, Z, A, W) = P(Y | M, Z, W)$ and $P(M | Z, A, W) = P(M | Z, W)$. We can factorize the likelihood for this SCM as follows: $P(O) = P(Y | M, Z, W)P(M|Z,W)P(Z | A, W)P(A | W)P(W)$.

The data-dependent, stochastic mediation estimand $E(Y_{a,\hat{g}_{M|a^*W}})$ can be identified via sequential regression, which provides the framework for our proposed estimator that follows. For intervention $(A=a, M=\hat{g}_{M|a^*,W})$, we have \\ \noindent $\bar{Q}_M^{\hat{g}}(Z,A,W)\equiv E_{\hat{g}_{M|a^*,W}}(E(Y |M,Z,A,W)|Z,A,W)$, where we integrate out $M$ under our stochastic intervention $\hat{g}_{M|a^*,W}$. This is accomplished by evaluating the inner expectation at each $m$ and multiplying it by the probability that $M=m$ under $\hat{g}_{M|a^*,W}$, summing over all $m$. We then integrate out $Z$ and set $A=a$: $\bar{Q}_Z^a(W)\equiv E_{Z}(\bar{Q}_M^{\hat{g}}(Z,A,W)|A=a,W)$. Taking the empirical mean gives the statistical parameter: $\Psi(P)(a,\hat{g}_{M|a^*,W})=E_W(\bar{Q}_Z^a(W)|W)$. 

The EIC for the parameter $\Psi(P)(a, \hat{g}_{M|a^*,W})$ is given by\begin{equation} \begin{split}
    D^*(a, \hat{g}_{M|a^*,W}) & = \sum_{k=0}^2 D_k^*(a, \hat{g}_{M|a^*,W}), \text{ where }\\
D^*_0(a, \hat{g}_{M|a^*,W}) & = \bar{Q}^a_Z(W) - \Psi(P)(a, \hat{g}_{M|a^*,W})\\
    D^*_1(a, \hat{g}_{M|a^*,W}) & = \frac{I(A=a)}{P(A=a | W)}(\bar{Q}^{\hat{g}}_M(Z,W) - \bar{Q}^a_Z(W))\\
    D^*_2(a, \hat{g}_{M|a^*,W}) & = \frac{I(A=a)\{I(M=1)\hat{g}_{M|a^*,W} + I(M=0)(1-\hat{g}_{M|a^*,W}) \}}{P(A=a | W)\{I(M=1)g_{M|Z,W} + I(M=0)(1-g_{M|Z,W}) \}}(Y-\bar{Q}_Y(M,Z,W)).   
\end{split}
\end{equation}

We now describe how to compute the TMLE. The estimation of $\hat{g}_{M|a^*,W}(W)$ does not differ from that described in the main text. 

\begin{enumerate}
\item Let $\bar{Q}_{Y,n}(M,Z,W)$ be an estimate of $\bar{Q}_Y(M,Z,W)\equiv E(Y |M,Z,W)$. To obtain $\bar{Q}_{Y,n}(M,Z,W)$, predict values of $Y$ from a regression of $Y$ on $M,Z,W$. 
\item Estimate the weights to be used for the initial targeting step:\\ $h_1(a) = \frac{I(A=a)\{I(M=1)\hat{g}_{M|a^*,W} + I(M=0)(1-\hat{g}_{M|a^*,W}) \}}{P(A=a)\{I(M=1)g_{M|Z,W} + I(M=0)(1-g_{M|Z,W}) \}},$ where $\hat{g}_{M|Z,W}$ are predicted probabilities from a logistic regression of $M=m$ on $Z$ and $W$. Let $h_{1,n}(a)$ denote the estimate of $h_1(a)$. 
\item Target the estimate of $\bar{Q}_{Y,n}(M,Z,W)$ by considering a univariate parametric submodel $\{\bar{Q}_{Y,n}(M,Z,W)(\epsilon):\epsilon\}$ defined as:
$logit(\bar{Q}_{Y,n} (\epsilon)(M,Z,W)) = logit(\bar{Q}_{Y,n}(M,Z,W) ) + \epsilon$. 
 Let $\epsilon_n$ be the MLE fit of $\epsilon$. We obtain $\epsilon_n$ by setting $\epsilon$ as the intercept of a weighted logistic regression model of $Y$ with $logit(\bar{Q}_{Y,n}(M,Z,W))$ as an offset and weights $h_{1,n}(a)$. (Note that this is just one possible TMLE.)
 The update is given by $ \bar{Q}^*_{Y,n}(M,Z,W) =  \bar{Q}_{Y,n}(\epsilon_n)(M,Z,W).$ $Y$ can be bounded to the [0,1] scale.
\item Let $\bar{Q}^{\hat{g}}_{M,n}(Z,W)$ be an estimate of $\bar{Q}^{\hat{g}}_{M}(Z,W)$. To obtain  $\bar{Q}^{\hat{g}}_{M,n}(Z,W)$, we integrate out $M$ from $\bar{Q}^*_{Y,n}(M,Z,W)$. First, we estimate $\bar{Q}^*_{Y,n}(M,Z,W)$ setting $m=1$ and $m=0$, giving $\bar{Q}^*_{Y,n}(m=1,Z,W)$ and $\bar{Q}^*_{Y,n}(m=0,Z,W)$. Then, multiply these predicted values by their probabilities under $\hat{g}_{M|a^*,W}(W)$ (for $a \in \{a,a^*\}$), and add them together (i.e., $\bar{Q}^{\hat{g}}_{M,n}(Z,W) = \bar{Q}^*_{Y,n}(m=1,Z,W)*\hat{g}_{M|a^*,W} + \bar{Q}^*_{Y,n}(m=0,Z,W)*(1-\hat{g}_{M|a^*,W})$). 
\item 
 We now fit a regression of $\bar{Q}^{\hat{g},*}_{M,n}(Z,W)$ on $W$ among those with $A=a$. We call the predicted values from this regression $\bar{Q}^{a}_{Z,n}(W)$. 
\item Complete a second targeting step: $logit(\bar{Q}^{a}_{Z,n}(\epsilon)(W)) = logit(\bar{Q}^{a}_{Z,n}(W)) + \epsilon h_{2,n}(a)$, where $h_{2,n}(a)$ is an estimate of $h_2(a) = \frac{I(A=a)}{P(A=a | W)}$ and $P(A=a | W)$ can be estimated from a logistic regression model of $A=a$ on $W$. Let $\epsilon_n$ again be the MLE fit of $\epsilon$, which can be obtained by fitting an intercept-only weighted logistic regression model of $\bar{Q}^{\hat{g}}_{M,n}(Z,W)$ with $logit(\bar{Q}^{a}_{Z,n}(W))$ as an offset and weights $h_{2,n}(a)$. (Alternatively, we could fit an unweighted logistic regression model of $\bar{Q}^{\hat{g}}_{M,n}(Z,W)$ with $logit(\bar{Q}^{a}_{Z,n}(W))$ as an offset and $h_{2,n}(a)$ as a covariate, where $\epsilon_n$ is the fitted coefficient on $h_{2,n}(a)$.) The update is given by $\bar{Q}^{a,*}_{Z,n}(W) = \bar{Q}^{a,a^*}_{Z,n}(\epsilon_n)(A,W).$
\item The TMLE of $\Psi(P)(a, \hat{g}_{M|a^*,W})$ is the empirical mean of $\bar{Q}^{a,*}_{Z,n}(W)$.
\item Repeat the above steps for each of the interventions. For example, for binary $A$, we would execute these steps a total of three times to estimate: 1) $\Psi(P)(1,\hat{g}_{M|1,W})$, 2) $\Psi(P)(1,\hat{g}_{M|0W})$, and 3) $\Psi(P)(0,\hat{g}_{M|0,W})$. 
\item The SDE can then be obtained by substituting estimates of parameters $\Psi(P)(a,\hat{g}_{M|a^*,W}) - \Psi(P)(a^*,\hat{g}_{M|a^*,W})$ and the SIE can be obtained by substituting estimates of parameters $\Psi(P)(a,\hat{g}_{M|a,W}) - \Psi(P)(a,\hat{g}_{M|a^*,W})$.
\item The variance of each estimate can be estimated as the sample variance of the EIC (defined above, substituting in the targeted fits $\bar{Q}^*_{Y,n}(M,Z,W)$ and $\bar{Q}^{a,*}_{Z,n}(W)$) divided by $n$. First, we estimate the EIC for each component of the SDE/SIE, which we call $EIC_{\Psi(P)(a,\hat{g}_{M|a^*,W})}$. 
 Then we estimate the EIC for the estimand of interest by subtracting the EICs corresponding to the components of the estimand. For example $EIC_{SDE} = EIC_{\Psi(P)(a,\hat{g}_{M|a^*,W})} - EIC_{\Psi(P)(a^*,\hat{g}_{M|a^*,W})}$. The sample variance of this EIC divided by $n$ is the influence curve-based variance of the estimator. 
\end{enumerate}

\subsection{There exist direct effects between $A$ and $M$ and between $A$ and $Y$}
Let observed data: $O=(W, A, Z, M, Y)$ with $n$ i.i.d. copies $O_1,...,O_n \sim P_0$, where $W$ is a vector of pre-treatment covariates, $A$ is the treatment, $Z$ is the intermediate confounder affected by $A$, $M$ is the mediator, and $Y$ is the outcome. For simplicity, we assume that $A, Z, M, \text{ and }Y$ are binary. We assume that $A$ is exogenous, $Z$ is affected by $\{A, W\}$, $M$ is affected by $\{A, Z, W\}$, and that $Y$ is affected by $\{A, M, Z, W\}$. We assume exogenous random errors: ($U_W, U_A, U_Z, U_M, U_Y$). We can factorize the likelihood for this SCM as follows: $P(O) = P(Y | M, Z, A,W)P(M|Z,A,W)P(Z | A, W)P(A)P(W)$.

Under this SCM, our proposed TMLE is efficient. The sequential regression used to identify the data-dependent, stochastic mediation estimands does not change from the that given in the main text for this SCM. 

The EIC for the parameter $\Psi(P)(a, \hat{g}_{M|a^*,W})$ is given by\begin{equation} \begin{split}
    D^*(a, \hat{g}_{M|a^*,W}) & = \sum_{k=0}^2 D_k^*(a, \hat{g}_{M|a^*,W}), \text{ where }\\
D^*_0(a, \hat{g}_{M|a^*,W}) & = \bar{Q}^a_Z(W) - \Psi(P)(a, \hat{g}_{M|a^*,W})\\
    D^*_1(a, \hat{g}_{M|a^*,W}) & = \frac{I(A=a)}{P(A=a)}(\bar{Q}^{\hat{g}}_M(Z,A,W) - \bar{Q}^a_Z(W))\\
    D^*_2(a, \hat{g}_{M|a^*,W}) & = \frac{I(A=a)\{I(M=1)\hat{g}_{M|a^*,W} + I(M=0)(1-\hat{g}_{M|a^*,W}) \}}{P(A=a)\{I(M=1)g_{M|Z,A,W} + I(M=0)(1-g_{M|Z,A,W}) \}}(Y-\bar{Q}_Y(M,Z,A,W)).   
\end{split}
\end{equation}

We now describe how to compute the TMLE. First, one estimates $\hat{g}_{M|a^*,W}(W) = \sum_{z=0}^1 P(M=1|Z=z,A=a^*,W)P(Z=z | A=a^*, W)$. Consider a binary $Z$. We first estimate $g_{Z|a^*,W}(W)=P(Z=1 | A=a^*, W)$. We then estimate $g_{M|z,a^*,W}(W)=P(M=1 | Z=z, A=a^*, W)$ for $z \in \{0,1\}$. We use these quantities to calculate $\hat{g}_{M|a^*,W} = (\hat{g}_{M|z=1,a^*,W} \times \hat{g}_{Z|a^*,W}) +  (\hat{g}_{M|z=0,a^*,W} \times (1-\hat{g}_{Z|a^*,W}))$. We can obtain $\hat{g}_{Z|a^*,W}(W)$ from a logistic regression of $Z$ on $A, W$ setting $A=a^*$, and $\hat{g}_{M|z,a^*,W}(W)$ from a logistic regression of $M$ on $Z, A, W$, setting $Z=\{0,1\}$ and $A=a^*$. We will then use this data-dependent stochastic intervention in the TMLE, whose implementation is described as follows.

\begin{enumerate}
\item Let $\bar{Q}_{Y,n}(M,Z,A,W)$ be an estimate of $\bar{Q}_Y(M,Z,A,W)\equiv E(Y |M,Z,A,W)$. To obtain $\bar{Q}_{Y,n}(M,Z,A,W)$, predict values of $Y$ from a regression of $Y$ on $M,Z,A,W$. 
\item Estimate the weights to be used for the initial targeting step:\\ $h_1(a) = \frac{I(A=a)\{I(M=1)\hat{g}_{M|a^*,W} + I(M=0)(1-\hat{g}_{M|a^*,W}) \}}{P(A=a)\{I(M=1)g_{M|Z,A,W} + I(M=0)(1-g_{M|Z,A,W}) \}},$ where $\hat{g}_{M|Z,A,W}$ are predicted probabilities from a logistic regression of $M=m$ on $Z$, $A$, and $W$. Let $h_{1,n}(a)$ denote the estimate of $h_1(a)$. 
\item Target the estimate of $\bar{Q}_{Y,n}(M,Z,A,W)$ by considering a univariate parametric submodel $\{\bar{Q}_{Y,n}(M,Z,A,W)(\epsilon):\epsilon\}$ defined as:
$logit(\bar{Q}_{Y,n} (\epsilon)(M,Z,A,W)) = logit(\bar{Q}_{Y,n}(M,Z,A,W) ) + \epsilon$. 
 Let $\epsilon_n$ be the MLE fit of $\epsilon$. We obtain $\epsilon_n$ by setting $\epsilon$ as the intercept of a weighted logistic regression model of $Y$ with $logit(\bar{Q}_{Y,n}(M,Z,A,W))$ as an offset and weights $h_{1,n}(a)$. (Note that this is just one possible TMLE.)
 The update is given by $ \bar{Q}^*_{Y,n}(M,Z,A,W) =  \bar{Q}_{Y,n}(\epsilon_n)(M,Z,A,W).$ $Y$ can be bounded to the [0,1] scale.
\item Let $\bar{Q}^{\hat{g}}_{M,n}(Z,A,W)$ be an estimate of $\bar{Q}^{\hat{g}}_{M}(Z,A,W)$. To obtain  $\bar{Q}^{\hat{g}}_{M,n}(Z,A,W)$, we integrate out $M$ from $\bar{Q}^*_{Y,n}(M,Z,A,W)$. First, we estimate $\bar{Q}^*_{Y,n}(M,Z,A,W)$ setting $m=1$ and $m=0$, giving $\bar{Q}^*_{Y,n}(m=1,Z,A,W)$ and $\bar{Q}^*_{Y,n}(m=0,Z,A,W)$. Then, multiply these predicted values by their probabilities under $\hat{g}_{M|a^*,W}(W)$ (for $a \in \{a,a^*\}$), and add them together (i.e., $\bar{Q}^{\hat{g}}_{M,n}(Z,A,W) = \bar{Q}^*_{Y,n}(m=1,Z,A,W)*\hat{g}_{M|a^*,W} + \bar{Q}^*_{Y,n}(m=0,Z,A,W)*(1-\hat{g}_{M|a^*,W})$). 
\item 
 We now fit a regression of $\bar{Q}^{\hat{g},*}_{M,n}(Z,A,W)$ on $W$ among those with $A=a$. We call the predicted values from this regression $\bar{Q}^{a}_{Z,n}(W)$. The empirical mean of these predicted values is the TMLE estimate of $\Psi(P)(a, \hat{g}_{M|a^*,W})$.
\item Repeat the above steps for each of the interventions. For example, for binary $A$, we would execute these steps a total of three times to estimate: 1) $\Psi(P)(1,\hat{g}_{M|1,W})$, 2) $\Psi(P)(1,\hat{g}_{M|0W})$, and 3) $\Psi(P)(0,\hat{g}_{M|0,W})$. 
\item The SDE can then be obtained by substituting estimates of parameters $\Psi(P)(a,\hat{g}_{M|a^*,W}) - \Psi(P)(a^*,\hat{g}_{M|a^*,W})$ and the SIE can be obtained by substituting estimates of parameters $\Psi(P)(a,\hat{g}_{M|a,W}) - \Psi(P)(a,\hat{g}_{M|a^*,W})$.
\item The variance of each estimate can be estimated as the sample variance of the EIC (defined above, substituting in the targeted fits $\bar{Q}^*_{Y,n}(M,Z,W)$ and $\bar{Q}^{a,*}_{Z,n}(W)$) divided by $n$. First, we estimate the EIC for each component of the SDE/SIE, which we call $EIC_{\Psi(P)(a,\hat{g}_{M|a^*,W})}$. 
 Then we estimate the EIC for the estimand of interest by subtracting the EICs corresponding to the components of the estimand. For example $EIC_{SDE} = EIC_{\Psi(P)(a,\hat{g}_{M|a^*,W})} - EIC_{\Psi(P)(a^*,\hat{g}_{M|a^*,W})}$. The sample variance of this EIC divided by $n$ is the influence curve-based variance of the estimator. 
\end{enumerate}

\section{Function code}
\begin{lstlisting}
medtmle<-function(a, z, m, y, w, svywt, zmodel, mmodel, ymodel, qmodel, gm, gma1){
  set.seed(34059)
  datw<-w
  tmpdat<-data.frame(cbind(datw, a=a))

  #get inital fit Q_Y
  tmpdat$qyinit<-cbind(predict(glm(formula=ymodel, family="binomial", data=data.frame(cbind(datw, z=z, m=m, y=y))), newdata=data.frame(cbind(datw, z=z, m=m)), type="response"), 
    predict(glm(formula=ymodel, family="binomial", data=data.frame(cbind(datw, z=z, m=m, y=y))), newdata=data.frame(cbind(datw, z=z, m=0)), type="response"),
    predict(glm(formula=ymodel, family="binomial", data=data.frame(cbind(datw, z=z, m=m, y=y))), newdata=data.frame(cbind(datw, z=z, m=1)), type="response"))
  
  #estimate weights for targeting
  psa1<-I(a==1)/mean(a)
  psa0<-I(a==0)/mean(1-a)
  mz<-predict(glm(formula=mmodel, family="binomial", data=data.frame(cbind(datw, z=z, m=m))), newdata=data.frame(cbind(datw, z=z)), type="response")
  psm<-(mz*m) + ((1-mz)*(1-m))
  
  tmpdat$ha1gma1<-((m*gma1 + (1-m)*(1-gma1))/psm) * psa1 * svywt
  tmpdat$ha1gma0<-((m*gm + (1-m)*(1-gm))/psm) * psa1 * svywt
  tmpdat$ha0gma0<-((m*gm + (1-m)*(1-gm))/psm) * psa0 * svywt
  
  tmpdat$y<-y

  #target Q_Y
  #for E(Y_{1,gmastar})
  epsilonma1g0<-coef(glm(y ~  1 , weights=tmpdat$ha1gma0, offset=(qlogis(qyinit[,1])), family="quasibinomial", data=tmpdat))
  tmpdat$qyupm0a1g0<-plogis(qlogis(tmpdat$qyinit[,2]) + epsilonma1g0)
  tmpdat$qyupm1a1g0<-plogis(qlogis(tmpdat$qyinit[,3]) + epsilonma1g0)
  #for E(Y_{1,gma})
  epsilonma1g1<-coef(glm(y ~  1 , weights=tmpdat$ha1gma1, offset=(qlogis(qyinit[,1])), family="quasibinomial", data=tmpdat))
  tmpdat$qyupm0a1g1<-plogis(qlogis(tmpdat$qyinit[,2]) + epsilonma1g1)
  tmpdat$qyupm1a1g1<-plogis(qlogis(tmpdat$qyinit[,3]) + epsilonma1g1)
  #for E(Y_{0,gmastar})
  epsilonma0g0<-coef(glm(y ~  1 , weights=tmpdat$ha0gma0, offset=(qlogis(qyinit[,1])), family="quasibinomial", data=tmpdat))
  tmpdat$qyupm0a0g0<-plogis(qlogis(tmpdat$qyinit[,2]) + epsilonma0g0)
  tmpdat$qyupm1a0g0<-plogis(qlogis(tmpdat$qyinit[,3]) + epsilonma0g0)

  #estimate Q_M
  tmpdat$Qma1g0<-tmpdat$qyupm0a1g0*(1-gm) + tmpdat$qyupm1a1g0*gm
  tmpdat$Qma1g1<-tmpdat$qyupm0a1g1*(1-gma1) + tmpdat$qyupm1a1g1*gma1
  tmpdat$Qma0g0<-tmpdat$qyupm0a0g0*(1-gm) + tmpdat$qyupm1a0g0*gm

  #estimate Q_Z
  Qzfita1g0<-glm(formula=paste("Qma1g0", qmodel, sep="~"), data=tmpdat[tmpdat$a==1,], family="quasibinomial")
  Qzfita1g1<-glm(formula=paste("Qma1g1", qmodel, sep="~"), data=tmpdat[tmpdat$a==1,], family="quasibinomial")
  Qzfita0g0<-glm(formula=paste("Qma0g0", qmodel, sep="~"), data=tmpdat[tmpdat$a==0,], family="quasibinomial")
  
  Qza1g0<-predict(Qzfita1g0, type="response", newdata=tmpdat)
  Qza1g1<-predict(Qzfita1g1, type="response", newdata=tmpdat)
  Qza0g0<-predict(Qzfita0g0, type="response", newdata=tmpdat)

  #update Q_Z 
  #Note: only need to do the update step if A is nonrandom
  epsilonza1g0<-coef(glm(Qma1g0~ 1 , weights=psa1*svywt, offset=qlogis(Qza1g0), family="quasibinomial", data=tmpdat))
  epsilonza1g1<-coef(glm(Qma1g1~ 1 , weights=psa1*svywt, offset=qlogis(Qza1g1), family="quasibinomial", data=tmpdat))
  epsilonza0g0<-coef(glm(Qma0g0~ 1 , weights=psa0*svywt, offset=qlogis(Qza0g0), family="quasibinomial", data=tmpdat))

  Qzupa1g0<-plogis(qlogis(Qza1g0) + epsilonza1g0)
  Qzupa1g1<-plogis(qlogis(Qza1g1) + epsilonza1g1)
  Qzupa0g0<-plogis(qlogis(Qza0g0) + epsilonza0g0)
  
  #estimate psi
  tmlea1m0<-sum(Qzupa1g0*svywt)/sum(svywt)
  tmlea1m1<-sum(Qzupa1g1*svywt)/sum(svywt)
  tmlea0m0<-sum(Qzupa0g0*svywt)/sum(svywt)

  #EIC
  tmpdat$qyupa1g0<-plogis(qlogis(tmpdat$qyinit[,1]) + epsilonma1g0)
  tmpdat$qyupa1g1<-plogis(qlogis(tmpdat$qyinit[,1]) + epsilonma1g1)
  tmpdat$qyupa0g0<-plogis(qlogis(tmpdat$qyinit[,1]) + epsilonma0g0)

  eic1a1g0<-tmpdat$ha1gma0 * (tmpdat$y - tmpdat$qyupa1g0)
  eic2a1g0<-psa1*svywt*(tmpdat$Qma1g0- Qzupa1g0)
  eic3a1g0<-Qzupa1g0-tmlea1m0
  eica1g0<-eic1a1g0 + eic2a1g0 + eic3a1g0

  eic1a1g1<-tmpdat$ha1gma1 * (tmpdat$y - tmpdat$qyupa1g1)
  eic2a1g1<-psa1*svywt*(tmpdat$Qma1g1- Qzupa1g1)
  eic3a1g1<-Qzupa1g1-tmlea1m1
  eica1g1<-eic1a1g1 + eic2a1g1 + eic3a1g1

  eic1a0g0<-tmpdat$ha0gma0 * (tmpdat$y - tmpdat$qyupa0g0)
  eic2a0g0<-psa0*svywt*(tmpdat$Qma0g0- Qzupa0g0)
  eic3a0g0<-Qzupa0g0-tmlea0m0
  eica0g0<-eic1a0g0 + eic2a0g0 + eic3a0g0

  #estimands
  nde<-tmlea1m0-tmlea0m0
  ndeeic<-eica1g0 - eica0g0
  vareic<-var(ndeeic)/nrow(tmpdat)

  nie<-tmlea1m1-tmlea1m0
  nieeic<-eica1g1 - eica1g0
  varnieeic<-var(nieeic)/nrow(tmpdat)

  return(list("nde"=nde, "ndevar"=vareic, "nie"=nie, "nievar"=varnieeic))
}
\end{lstlisting}
\begin{lstlisting}
set.seed(2350) 

n<-100000

w0<-rbinom(n, 1, .5)
w1<-rbinom(n,1, .4 + (.2*w0))

probsel<-plogis(-1+ log(4)*w1 + log(4)*w0)
psel<-rbinom(n, 1, probsel)
svywt<-mean(probsel)/probsel

#instrument
a<-rbinom(n, 1, .5)
  
#exposure
z0<-rbinom(n,1,plogis(       - log(2)*w1))
z1<-rbinom(n,1,plogis(log(4) - log(2)*w1))
z<-ifelse(a==1, z1, z0)

#mediator
m0<-rbinom(n, 1, plogis(-log(3)          - log(1.4)*w1))
m1<-rbinom(n, 1, plogis(-log(3) + log(10)- log(1.4)*w1))
m<-ifelse(z==1, m1, m0)

#outcomes
y<-rbinom(n,1, plogis(log(1.2)  + (log(3)*z)  + log(3)*m - log(1.2)*w1 + log(1.2)*w1*z) )

dat<-data.frame(w1=w1, a=a, z=z, m=m, y=y, psel=psel, svywt=svywt, radid_person=seq(1,n,1))
obsdat<-dat[dat$psel==1,]
 
zmodel<-"z ~ a + w1 "
mmodel<-"m ~ z + w1"
ymodel<-"y ~ m + z*w1"
qmodel<-"w1"

#make gm
zfit<-glm(formula=zmodel, family="binomial", data=obsdat)
mfit<-glm(formula=mmodel, family="binomial", data=obsdat)

za0<-predict(zfit, newdata=data.frame(w1=obsdat$w1, a=0), type="response")
za1<-predict(zfit, newdata=data.frame(w1=obsdat$w1, a=1), type="response")

mz1<-predict(mfit, newdata=data.frame(w1=obsdat$w1, z=1), type="response")
mz0<-predict(mfit, newdata=data.frame(w1=obsdat$w1, z=0), type="response")

gm<-(mz1*za0) + (mz0*(1-za0))
gma1<-(mz1*za1) + (mz0*(1-za1))

res<-medtmle(a=obsdat$a, z=obsdat$z, m=obsdat$m, y=obsdat$y, w=data.frame(w1=obsdat$w1), svywt=obsdat$svywt, zmodel=zmodel, mmodel=mmodel, ymodel=ymodel, qmodel=qmodel, gm=gm, gma1=gma1)
\end{lstlisting}

\end{document}